\documentclass[aps,prb,article,twocolumn,preprintnumbers,amsmath,amssymb,superscriptaddress]{revtex4}

\date{\today}
\usepackage{epsfig}
\usepackage{subfigure}
\usepackage{graphicx}
\usepackage{dcolumn}
\usepackage{bm}
\usepackage[colorlinks,linkcolor=blue,hyperindex,CJKbookmarks]{hyperref}
\usepackage{float}
\usepackage{hyperref}
\usepackage{comment}

\newcommand{\tJ}{$t\small{-}J$}
\newcommand{\tJs}{$t\small{-}J\small{-} 3s$}

\newcommand{\Tr}    {{\mathrm{Tr}}}
\newcommand{\Ham}   {{\mathcal{H}}}

\newcommand{\ibf}      {\textbf{i}}
\newcommand{\jbf}      {\textbf{j}}
\newcommand{\rbf}      {\textbf{r}}

\begin{document}

\title{Higher-Order Spin-Hole Correlations around a Localized Charge Impurity}
\author{Yao Wang}
\email{yaowang@g.clemson.edu}
\affiliation{Department of Physics and Astronomy, Clemson University, Clemson, South Carolina 29631, USA}
\author{Annabelle Bohrdt}
\affiliation{Department of Physics and Institute for Advanced Study, Technical University of Munich, 85748 Garching, Germany}
\affiliation{Munich Center for Quantum Science and Technology (MCQST), 80799 M\"unchen, Germany}
\author{Joannis Koepsell}
\affiliation{Max-Planck-Institut f\"ur Quantenoptik, 85748 Garching, Germany}
\affiliation{Munich Center for Quantum Science and Technology (MCQST), 80799 M\"unchen, Germany}
\author{Eugene Demler}
\affiliation{Department of Physics, Harvard University, Cambridge 02138, USA}
\author{Fabian Grusdt}
\email{Fabian.Grusdt@physik.uni-muenchen.de}
\affiliation{Department of Physics and Arnold Sommerfeld Center for Theoretical Physics (ASC), Ludwig-Maximilians-Universit\"at M\"unchen, Theresienstr. 37, M\"unchen D-80333, Germany}
\affiliation{Munich Center for Quantum Science and Technology (MCQST), 80799 M\"unchen, Germany}

\date{\today}
\begin{abstract}
Analysis of higher order correlation functions has become a powerful tool for investigating interacting many-body systems in quantum simulators, such as quantum gas microscopes. Experimental measurements of mixed spin-charge correlation functions in the 2D Hubbard have been used to study equilibrium properties of magnetic polarons and to identify coherent and incoherent regimes of their dynamics. In this paper we consider theoretically an extension of this technique to systems which use a pinning potential to reduce the mobility of a single dopant in the Mott insulating regime of the 2D Hubbard model. We find that localization of the dopant has a dramatic effect on its magnetic dressing. The connected third order spin correlations are weakened in the case of a mobile hole but strengthened near an immobile hole. In the case of the fifth-order correlation function, we find that its bare value has opposite signs in cases of the mobile and of fully pinned dopant, whereas the connected part is similar for both cases. We study suppression of higher order correlators by thermal fluctuations and demonstrate that they can be observed up to temperatures comparable to the spin-exchange energy $J$. We discuss implications of our results for understanding the interplay of spin and charge in doped Mott insulators.
\end{abstract}
\pacs{xxx }

\maketitle

\section{Introduction}
Understanding strongly correlated systems and their emergent phases of matter has been a central task in modern condensed matter physics. Cuprate superconductors have provided a particularly strong motivation for this line of inquiry because of both potential applications of high-temperature superconductivity and unusual thermodynamic and transport properties of these materials\,\cite{dagotto1994correlated}. The single-band (Fermi) Hubbard model has been commonly considered as the minimal microscopic model for describing the physics of high-$T_c$ curates\,\cite{anderson1987resonating,zhang1988effective,lee2006doping}. It correctly reproduces the antiferromagnetic (AFM) Mott insulating state at half filling\,\cite{singh2002spin}, and is believed to exhibit non-Fermi liquid properties\,\cite{zheng2017stripe, huang2017numerical,ponsioen2019period, kokalj2017bad, huang2019strange, cha2020slope} as well as $d$-wave pairing at finite doping\,\cite{maier2005systematic, zheng2016ground,ido2018competition,jiang2019superconductivity,chung2020plaquette}, though other lattice and material-specific factors beyond the Hubbard model were also argued to be important\cite{ohta1991apex, lanzara2001evidence, shen2004missing, mishchenko2004electron, cuk2004coupling, reznik2006electron, johnston2012evidence, peng2017influence, he2018rapid}. 
Methods based on the mean-field approximation are not applicable for analyzing the Hubbard model, because of the presence of many competing instabilities and the importance of quantum antiferromagnetic fluctuations that underlie the emergent non-local attraction between electrons\cite{demler2004so,scalapino2012common,maier2016pairing,leblanc2015solutions}.

Theoretically, the motion of a single hole in a doped Hubbard model takes the form of spin-polaron propagation (sometimes also referred to as magnetic polaron), where the dopant is dressed by a cloud of spin defects in the AFM background.\cite{Kane1989,Sachdev1989,Liu1991,martinez1991spin, stephan1991fermi, manousakis1991spin, dagotto1992static, liu1992dynamical,stephan1992single, bala1995spin,Brunner2000} Fluctuations of these surrounding spins, correlated with the dopants, should also play a crucial role in the pairing of electrons or holes.\cite{scalapino1986d,gros1987superconducting,kotliar1988superexchange,schrieffer1989dynamic, scalapino1995case,tsuei2000pairing} This mechanism has been supported by multiple solid-state experiments. For example, photoemission (ARPES) experiments have revealed the presence of strong correlations, such as the ``high-energy anomaly'' and the non-quasiparticle features, even in overdoped cuprates\cite{graf2007universal, he2019fermi, chen2019incoherent}; recently resonant inelastic x-ray scattering experiments (RIXS) have also observed the persistent spin fluctuations in a wide range of electron and hole doping\cite{le2011intense, dean2013persistence, dean2013high, lee2014asymmetry, ishii2014high}. Therefore, the correlations between spin and dopants generally exist in strongly correlated materials and may be crucial for the observed emergent phases, including also the spin-glass phase at low doping in disordered cuprate compounds\,\cite{hasselmann2004spin}.

An additional line of inquiry into the nature of strongly correlated electron systems such as cuprates comes from using impurities.
One of the most well-studied examples is the non-magnetic Zn-substitution into the CuO$_2$ plane, which induces charge impurities due to local chemical potential shifts. With this substitution, nuclear magnetic resonance (NMR) and muon spin relaxation ($\mu$SR) experiments found that the magnetic moment is locally enhanced near the Zn impurities\,\cite{mahajan199489, alloul1991correlations,xiao1990magnetic,mendels1994muon}. At the same time, the substitution heavily suppresses the $d$-wave superconductivity\,\cite{xiao1990magnetic, maeda1990substitution, mendels1994muon, fukuzumi1996universal, nachumi1996muon,basov1998infrared}. While the mean-field theory can address this pair-breaking phenomenon\,\cite{lee1993localized,sun1995impurity,salkola1996theory, hirschfeld1996theory,franz1996impurity,tsuchiura1999quasiparticle}, it appears to provide an apparent contradiction to the spin-fluctuation-mediated pairing mechanism: It remains unclear why superconductivity is suppressed while the local moment increases.

As an insightful perspective, we know that it is the dynamical instead of static spin correlations that enhance the $d$-wave superconductivity\,\cite{kivelson1998electronic,emery1999stripe,kivelson2003detect}.
Therefore, to address this question, one needs to understand the microscopic correlation between the carrier (electron or hole) and the spin fluctuations near the impurity. This is important not only for cuprates but also generally for all correlated materials: As we will show in this paper, a local charge impurity has outsized effects on high-order spin correlations. However, such a correlation involves a dopant hole/electron and at least two neighboring spins, which cannot be directly accessed in existing solid-state spectral measurements. Fortunately, this task was recently achieved by a distinct approach -- the ultracold-atom based quantum simulator\,\cite{gross2017quantum}. With the help of quantum gas microscopes, the direct measurement of instantaneous correlations, especially those high-order correlations, becomes possible\cite{bakr2009quantum, sherson2010single, parsons2015site, cheuk2015quantum, omran2015microscopic, edge2015imaging, haller2015single,mazurenko2017cold, hilker2017revealing, chiu2019string, schweigler2017experimental}. The full spatial resolution of individual lattice sites has also been achieved. Furthermore, quantum simulators have control over dimensionality\,\cite{salomon2019direct} and can engineer tailored optical potentials to precisely control model Hamiltonians\,\cite{brown2019bad,nichols2019spin}. As an example, an optical tweezer can be used to engineer a localized potential for holes, which allows for continuous control of a local impurity in a clean system\cite{koepsell2019imaging}. This paradigmatic scenario of a single disordered lattice site with a tunable on-site potential will be analyzed in detail throughout this work.

Here, to systematically understand the high-order spin-hole correlations in the doped Hubbard model and the impact of an impurity on the correlations, we present an exact diagonalization calculation of \tJs\ and Hubbard models. Compared to the well-known spin polaron dressing in the case of a mobile dopant in a Mott insulator, we find that the presence of a local pinning field of variable strength results in significantly different spin correlation patterns. Such a difference is reflected in both third-order and fifth-order spin-hole correlation functions, suggesting a crossover from a spin-polaron surrounded by weakened magnetic correlations to a geometric defect surrounded by enhanced magnetic fluctuations. This work also complements our analysis of the fifth-order correlations of unpinned holes in Ref.~\onlinecite{bohrdt2020shortPaper}. Since the proposed correlation functions are composed of local spin and charge operators, they can be measured directly using the state-of-the-art quantum gas microscopy\,\cite{koepsell2019imaging, koepsell2020microscopic}, which can ultimately extend the quantitative conclusion to the thermodynamic limit.

This paper is organized as follows. We first introduce the models and methods used in Sec.~\ref{sec:modelMethod}. Then we investigate the third-order and the fifth-order correlation functions in Sec.~\ref{sec:3rdOrder} and Sec.~\ref{sec:5thOrder}, respectively. We conclude our study in Sec.~\ref{sec:conclusion}.

\section{Models and Methods}\label{sec:modelMethod}
As the simplest microscopic model depicting Mott physics in a strongly correlated electronic system, the 2D (Fermi) Hubbard model is described by the Hamiltonian
\begin{eqnarray} \label{eq:hubbard}
\mathcal{H}_{H}&=&-\sum_{{\bf i},{\bf j},\sigma}  \left(t_{\bf ij} \hat{c}^\dagger_{{\bf j} \sigma} \hat{c}_{{\bf i} \sigma} +h.c.\right) \nonumber\\
 &&+ U\sum_{{\bf i}}\left(\!\hat{n}_{{\bf i}\uparrow}-\frac12\right)\left(\hat{n}_{{\bf i}\downarrow}-\frac12\right),
\end{eqnarray}
where $\hat{c}^\dagger_{{\bf i}\sigma}$ ($\hat{c}_{{\bf i}\sigma}$) and $\hat{n}_{{\bf i}\sigma}$ denotes the creation (annihilation) and density operator at site $\textbf{i}$ of spin $\sigma$; $t_{\bf ij}$ is the hopping, restricted here to nearest-neighbors (nn) $t_{\bf \langle ij\rangle}\!=\!t$. As mentioned above, the single-band Hubbard model is believed to capture the essential physics of strongly correlated materials such as cuprates \cite{anderson1987resonating,zhang1988effective} and can be precisely simulated by the cold-atom experiments\cite{hofstetter2002high, ho2009quantum, jordens2010quantitative, esslinger2010fermi, mazurenko2017cold, hilker2017revealing, tarruell2018quantum, chiu2019string}.

The Hubbard model's Hilbert space dimension is relatively large, limiting both the system size and the number of states accessible in exact numerical calculations. To investigate the temperature dependence in the regime accessible by cold-atom experiments (typically $T\sim 0.5t$), we also consider the low-energy approximation of the Hubbard model near half-filling -- the \tJ-like spin model. Through a $t/U$ expansion to the lowest order, one can simplify the Hubbard model to the \tJ\ model~\cite{chao1977tJmodel, chao1978canonical, belinicher1994consistent, belinicher1994range}.
\begin{equation}\label{eq:tJHam}
\mathcal{H}_{t\!-\!J}=-t\!\sum_{\langle {\bf i},{\bf j} \rangle,\sigma}\! \left( \tilde{c}^\dagger_{{\bf j} \sigma} \tilde{c}_{{\bf i}\sigma}\! +\!h.c.\right)\!+\!J\sum_{\langle {\bf i},{\bf j}\rangle}\left( \textbf{S}_{\bf i} \! \cdot \! \textbf{S}_{\bf j}-\frac{\hat{n}_\ibf \hat{n}_\jbf}{4}\right),
\end{equation}
where $\textbf{S}_{\bf i}\cdot\textbf{S}_{\bf j}\! =\! \hat{S}_{\bf i}^z\hat{S}^z_{\bf j}\!+\!\frac12\left(\hat{S}_{\bf i}^+\hat{S}_{\bf j}^-+\hat{S}_{\bf i}^-\hat{S}_{\bf j}^+\right)$, with $\hat{S}_{\bf i}^z\!=\!(\hat{n}_{{\bf i} \uparrow}\!-\!\hat{n}_{{\bf i}\downarrow})/2$ and $\hat{S}_{\bf i}^+\!=\!(\hat{S}_{\bf i}^-)^\dagger\!=\!\tilde{c}^\dagger_{{\bf i}\uparrow} \tilde{c}_{{\bf i}\downarrow}$.
The constrained fermionic operators acting in the Hilbert space without double occupancy are defined as $\tilde{c}^{\dag}_{{\bf i}\sigma} = \hat{c}_{{\bf i}\sigma}^{\dag}(1-\hat{n}_{{\bf i} \bar{\sigma}})$. Although sometimes negligible, there is a 3-site term at the same lowest order ($\sim t^2/U$) that contributes to the dopant's motion\,\cite{stephan1992single, jefferson1992derivation, bala1995spin, Spalek1988, Szczepanski1990, Eskes1994, Eskes1994b, belinicher1994consistent, belinicher1994range, belinicher1996generalized, belinicher1996single, psaltakis1992optical, Eskes1996}. This defines the \tJs\ model with the Hamiltonian given by $\mathcal{H}_{t\!-\!J\!-\!3s}= \mathcal{H}_{t \!-\!J}+\mathcal{H}_{3s}$:
\begin{equation}\label{tJ3sHam}
\mathcal{H}_{3s}= - \frac{J}{4}\!\!\sum_{\langle {\bf i},{\bf j}\rangle,\langle {\bf i},{\bf j}^\prime\rangle\atop {\bf j}\neq {\bf j}^\prime,\sigma}\! \! \left(\tilde{c}^\dagger_{{\bf j}^\prime\sigma}\tilde{n}_{{\bf i}\bar{\sigma}}\tilde{c}_{{\bf j}\sigma}\!+ \! \tilde{c}^\dagger_{{\bf j}^\prime\sigma} \tilde{c}^\dagger_{{\bf i}\bar{\sigma}} \tilde{c}_{{\bf i}\sigma} \tilde{c}_{{\bf j}\bar{\sigma}}\right).
\end{equation}
We use both the Hubbard and \tJs\ models to study high-order correlations, where the latter allows for access to higher temperatures. Comparisons between the two models allow identifying which effects are captured by effective spin models and do not require higher-order expansions in $t/U$. Unless otherwise specified, we use $U=8t$ for the Hubbard model and accordingly $J=0.5t$ for the \tJs\ model throughout this paper.

To test the correlations with respect to an immobile hole, we also consider an extra pinning potential in the origin (${\bf r}_0 = \bf 0$), controlling the mobility of the single hole. The Hamiltonian becomes:
\begin{equation}\label{pinnedHam}
\mathcal{H}= \mathcal{H}_0 + V \sum_{\sigma} \hat{n}_{{\bf 0}\sigma}\,
\end{equation}
where $\mathcal{H}_0$ is the Hubbard or \tJs\ Hamiltonian, and $V$ is the strength of a local pinning potential. This pinning potential can be realized experimentally by an optical tweezer in an ultracold atom system\,\cite{koepsell2019imaging}. Hence, the high-order correlations analyzed in this paper and Ref.~\onlinecite{bohrdt2020shortPaper} are directly accessible to these experiments. Note that the strength of spin-exchange and 3-site terms in the \tJ\ model involving a virtually doubly-occupied central site (which is perturbed by the added pinning potential) must be modified when $|V|$ becomes comparable to the Hubbard interaction $U$. Throughout this work, we neglect such corrections and focus on the regime $|V| \leq U$. The difference between Hubbard and \tJs\ models are quantitatively compared and discussed whenever relevant.

To resolve the wavefunction and the corresponding correlators over a range of low temperatures in a 2D $D_4$ symmetric system, we perform exact diagonalization calculations on a 4$\times$4 cluster with periodic boundary conditions. Throughout this paper, we evaluate the expectation values of observables in a canonical ensemble. For convenience, we denote the expectation as the thermal average
\begin{equation}
    \left\langle \hat{O}\right\rangle = \Tr\left[ \frac{e^{- \Ham/T}}{\mathcal{Z}} \hat{O} \right] \approx \sum_{n<n_{\rm max}}  \frac{e^{-E_n/T}}{\mathcal{Z}}\langle n|\hat{O}|n\rangle\,,
\end{equation}
in which $\mathcal{Z}$ is the partition function. The $n_{\rm max}$ sets the numerical truncation of excited states, which satisfies $E_{n_{\rm max}}-E_0 \gg T$ for all temperatures considered in our work.
In the results presented in Secs.~\ref{sec:3rdOrder} and \ref{sec:5thOrder}, we keep $n_{\rm max}\sim650$ states for the Hubbard model, giving reliable results up to $T\sim 0.4t$, whereas $n_{\rm max}\sim13,000$ states for the \tJs\ model, giving reliable results up to $T\sim t$.

In this paper, we employ the parallel Arnoldi and Paradeisos algorithm to determine the equilibrium ground-state wavefunction and expectation values.\cite{lehoucq1998arpack,jia2017paradeisos} Unless otherwise indicated, we include all total-$S^z$ sectors in the thermal ensemble. Part of the results in Sec.~\ref{sec:5thOrder} are benchmarked using the density matrix renormalization group (DMRG) at zero temperature in a $6\times12$ cluster with cylindrical boundary condition, to investigate possible artifacts caused by the finite system size.

\section{The Third-Order Spin-Hole Correlations}\label{sec:3rdOrder}

\begin{figure}[!th]
\begin{center}
\includegraphics[width=\columnwidth]{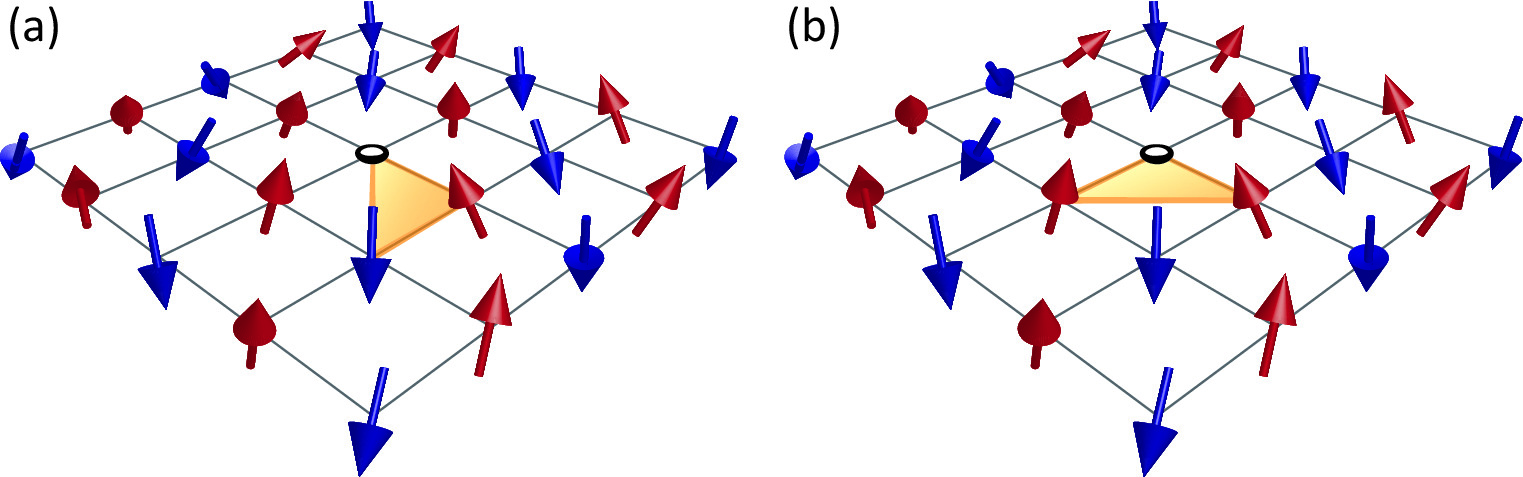}
\caption{\label{fig:3rdCartoon} Schematic showing the third-order spin-hole correlators for (a) nearest spins ($|\mathbf{d}| = 1$) and (b) next-nearest spins ($|\mathbf{d}| = \sqrt{2}$) .
}
\end{center}
\end{figure}

To describe the spin correlations with respect to a doped hole, we consider the following third-order correlation function: 
\begin{equation}
    B(\rbf, \rbf^\prime, \mathbf{d}) = 4\left\langle  \hat{n}^{h}_{\rbf} \hat{S}^z_{\rbf^\prime} \hat{S}^z_{\rbf^\prime+\mathbf{d}} \right\rangle/\langle \hat{n}_{\rbf}^{h}\rangle
\end{equation}
where the hole density operator $\hat{n}_{\rbf}^{h} = (1-\hat{n}_{\rbf\uparrow})(1-\hat{n}_{\rbf\downarrow})$. In this paper, we focus on the $|\mathbf{d}| = 1$ (neighboring spins) and $|\mathbf{d}| = \sqrt{2}$ (diagonal spins), allowing the relative distance between the hole and the spin ``bond'' $|\rbf-\rbf^\prime|$ to vary (see Fig.~\ref{fig:3rdCartoon}). 

\begin{figure}[!t]
\begin{center}
\includegraphics[width=\columnwidth]{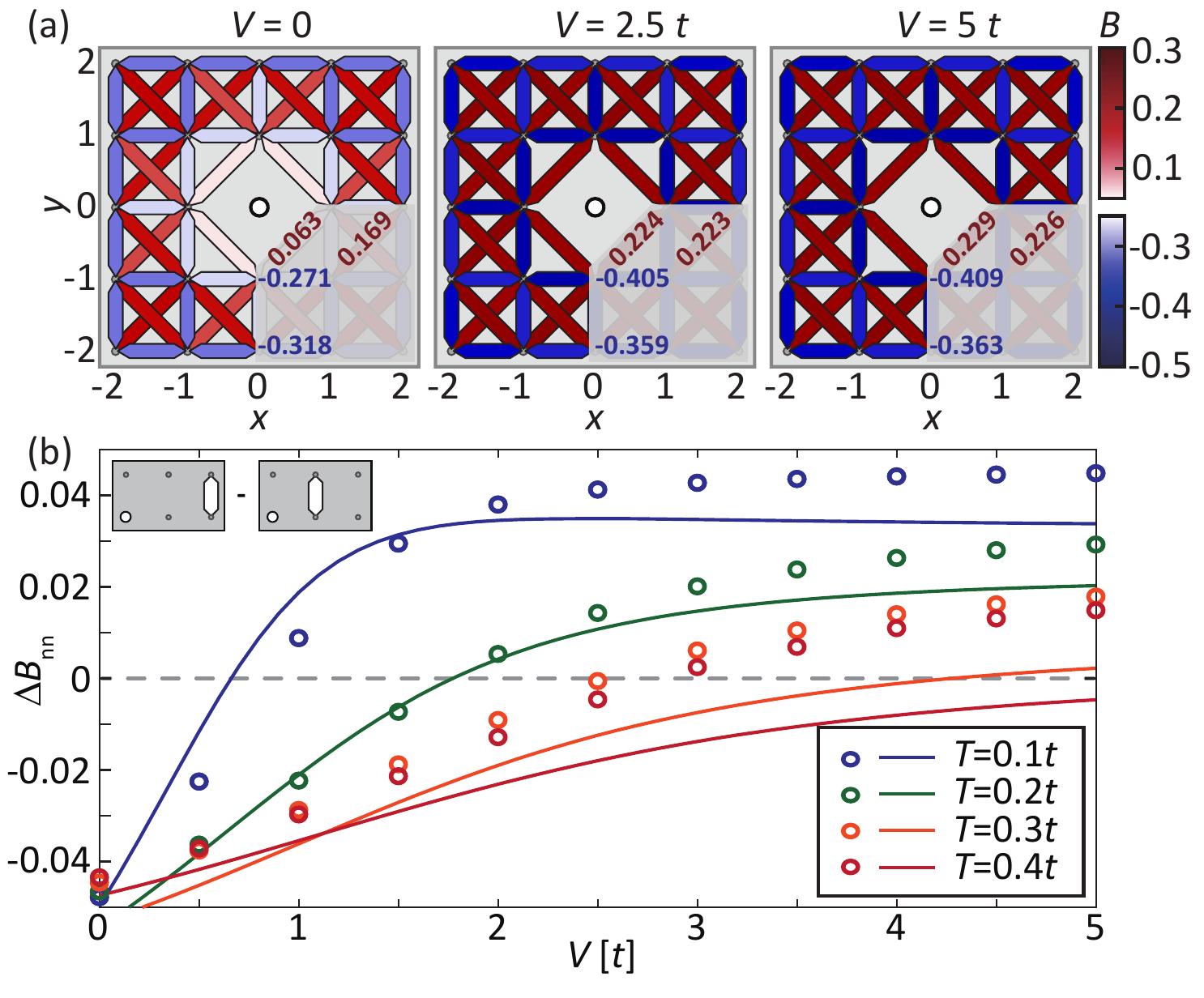}
\caption{\label{fig:3rdDifferential} (a) The spatial distribution of $B(\rbf, \rbf^\prime, \mathbf{d})$ for $|\mathbf{d}| = 1$ and $|\mathbf{d}| = \sqrt{2}$ in a $n=1/16$ doped Hubbard model at $T=0$, with pinning potential $V=0$, $V=2.5t$ and  $V=5t$, respectively. The numbers in the bottom-right corner of each panel mark the values of the correlators mentioned in the main text. (b) The $V$-dependence of the differential third-order correlators $\Delta B_{\rm nn}$ for $T=0.1t$ (blue), 0.2$t$ (green), 0.3$t$ (orange), and 0.4$t$ (red). The dots and solid lines denote the results obtained for the Hubbard and \tJs\ model, respectively. The inset in (b) explains the definition of the differential correlators.
}
\end{center}
\end{figure}

Figure \ref{fig:3rdDifferential} gives an overview of the spatial $(\rbf^\prime-\rbf)$ distribution of the $B(\rbf, \rbf^\prime, \mathbf{d})$ in a single-hole-doped Hubbard model at zero temperature, with $\rbf$ corresponding to the coordinate where the pinning potential $V$ is applied (white dot). For different pinning potentials, the nearest-neighbor ($|\mathbf{d}| = 1$) correlations are always negative, while the diagonal ($|\mathbf{d}| = \sqrt{2}$) correlations are mostly positive due to the strong AFM order. However, the spatial variation of these correlators indicates different underlying physics triggered by the pinning potential, as discussed below.

We first consider systems with a mobile hole, i.e., for $V=0$. As shown in the left panel of Fig.~\ref{fig:3rdDifferential}(a), the (absolute value of the) third-order correlator $B(\rbf, \rbf^\prime, \mathbf{d})$ is weakened for shorter distances $|\rbf-\rbf^\prime|$. This spatial distribution can be understood in terms of the spin polaron \cite{grusdt2019microscopic,blomquist2019ab}, where a dopant's motion is dressed by a cloud of spin defects, forming a polaronic quasiparticle. As the system is translationally invariant for $V=0$, the correlator $B(\rbf, \rbf^\prime, \mathbf{d})$ depicts the concentration of spin defects in the co-moving frame of the mobile hole. Due to the small radius of the spin polaron (local screening of spin correlations), the amplitudes of correlators for large $|\rbf^\prime-\rbf|$'s asymptotically approach those of an undoped AFM system.

\begin{figure*}[!th]
\begin{center}
\includegraphics[width=17cm]{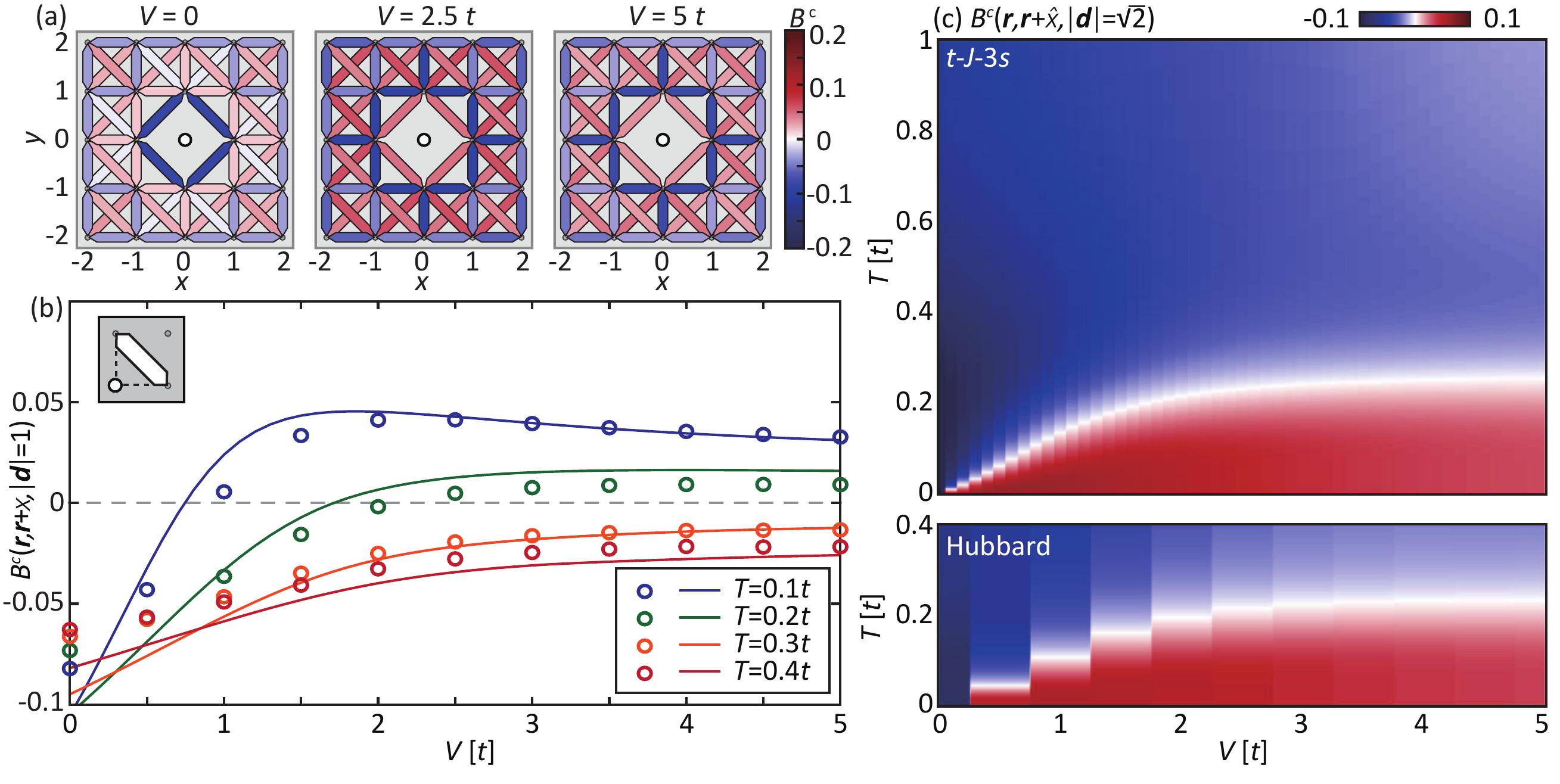}
\caption{\label{fig:3rdConnected} 
(a) The spatial distribution of connected parts $B^c(\rbf, \rbf^\prime, \mathbf{d})$ in a $n=1/16$ doped Hubbard model at $T=0$, with pinning potential $V=0$, $V=2.5t$ and $V=5t$, respectively.
(b) The $V$-dependence of the connected third-order correlators $B^c(\rbf, \rbf^\prime, \mathbf{d})$ for $T=0.1t$ (blue), 0.2$t$ (green), 0.3$t$ (orange), and 0.4$t$ (red). The dots and solid lines denote the results obtained for the Hubbard and \tJs\ model, respectively. The inset in (b) explains the definition of the differential correlators. (c) The temperature and pinning potential dependence of the genuine third-order correlator $B^c(\rbf, \rbf+\hat{x}, \mathbf{d})$ with $|\mathbf{d}|=\sqrt{2}$.
}
\end{center}
\end{figure*}

In contrast to the free hole, the spin correlations in the presence of a pinning potential $V > 0$ lead to significantly different patterns. With the pinning potential applied at site $\rbf$, we examine the the third-order correlator $B(\rbf, \rbf^\prime, \mathbf{d})$ between the hole (at $\rbf$) and the spins at $\rbf^\prime$ and $\rbf^\prime+\mathbf{d}$. As shown in the right two panels of Fig.~\ref{fig:3rdDifferential}(a) and Fig.~\ref{fig:3rdDifferential}(b), $B(\rbf, \rbf^\prime, \mathbf{d})$ becomes stronger for shorter distances $|\rbf-\rbf^\prime|$ (i.e., for $|\mathbf{d}|=1$ it becomes more negative, while for $|\mathbf{d}|=\sqrt{2}$ it becomes more positive). To distinguish from the screening of correlations in the spin-polaron picture, we denote this phenomenon as "anti-screening''. A similar phenomena has recently been observed in experiments, albeit at higher temperatures, where it has been attributed to imperfections of the pinning potential \cite{koepsell2019imaging}. Indirectly, this phenomenon is also consistent with the increased local moments measured in NMR and $\mu$SR experiments in Zn-substituted cuprates\,\cite{mahajan199489, alloul1991correlations,xiao1990magnetic,mendels1994muon}. One can intuitively understand this phenomenon by observing the fact that spin correlations in systems with a lower coordination numbers (e.g., a 1D chain) are stronger than in the 2D plane with otherwise identical parameters due to the existence of reduced frustrations (an AFM state favors spin to form singlets with all its nearest neighbors)\,\cite{essler1991complete, seabra2014real}. Thus, a system with a strong pinning potential tends to form a 1D edge at the boundary surrounding the impurity. Here the system can lower its energy by forming stronger singlet bonds closer to the impurity, while retaining the bulk spin order at further-away sites.
In addition to the change in spatial distribution, the presence of a pinning potential also leads to an overall enhancement of all correlators (in terms of absolute values). 

To quantify the different distributions of the third-order correlators, we define the difference between the nearest-neighbor correlators as [see the illustration in the inset of Fig.~\ref{fig:3rdDifferential}(b)]
\begin{equation}\label{eq:3rdDiffNN}
    \Delta B_{\rm nn} = B(\rbf, \rbf +2\hat{x}, \hat{y}) - B(\rbf, \rbf +\hat{x}, \hat{y}). 
\end{equation}
Through the difference, the uniform (positive or negative) background in the ``bare'' correlator $B(\rbf, \rbf^\prime, \mathbf{d})$ is removed. Thus, we refer to the $\Delta B_{\rm nn}$ as \textit{differential third-order correlators}. As shown in Fig.~\ref{fig:3rdDifferential}(b), $\Delta B_{\rm nn}$ is always negative for $V=0$, indicating the screening effect of the spin polaron; the $\Delta B_{\rm nn}$ turns to positive for a finite $V$, indicating the onset of anti-screening\cite{degeneracy}. As such, the crossover between screening and anti-screening natures is clearly reflected by the differential third-order correlators, while the original $B(\rbf, \rbf^\prime, \hat{y})$s do not have a sign change. The reason is that the original ones contain a substantial contribution from the lower-order correlators (i.e., $\langle  \hat{S}^z_{\rbf} \hat{S}^z_{\rbf+\hat{y}} \rangle$), which is sizably negative. A spatial difference with respect to the hole, however, cancels (or at least heavily reduces) this lower-order background.

The above observation indicates that a ``genuine'' correlator is required while describing the underlying many-body physics. Such a \textit{genuine third-order correlator} can be more intuitively defined as the connected part of $B(\rbf, \rbf^\prime, \mathbf{d})$
\begin{equation}
    B^{c}(\rbf, \rbf^\prime, \mathbf{d}) = B(\rbf, \rbf^\prime, \mathbf{d}) -  \sum_{\rbf^{\prime\prime}}\left\langle  \hat{S}^z_{\rbf^{\prime\prime}} \hat{S}^z_{\rbf^{\prime\prime}+\mathbf{d}} \right\rangle/N,
\end{equation}
where $N$ denotes the number of lattice sites. As such, the $B^{c}(\rbf, \rbf^\prime, \mathbf{d})$ reflects the net hole-spin-spin correlation on top of the AFM background, without the need to extract through a spatial difference. As shown in Fig.~\ref{fig:3rdConnected}(a), the $B^{c}(\rbf, \rbf^\prime, \mathbf{d})$'s for small $|\rbf-\rbf^\prime|$ flip sign in the presence of the pinning potential, which allows for a better discrimination of the screening and anti-screening regimes.

Given that thermal fluctuations are expected to disrupt long-range ordering, the crossover between the screening and anti-screening situations should depend on both temperature and the pinning potential. To address this, we calculate the $T$- and $V$-dependence of $B^{c}(\rbf, \rbf^\prime, \mathbf{d})$. As mentioned in Sec.~\ref{sec:modelMethod}, the Hilbert space size of the Hubbard model limits the accessible maximal temperature; therefore, we also calculate the above correlators for the \tJs\ model. As shown in Figs.~\ref{fig:3rdConnected}(b) and (c), the $\Delta B_{\rm nn}$ characterizes a crossover from a screening (negative) to the anti-screening (positive) regime for any temperature $T<0.3t$. The critical $V$ value increases with temperature, because thermal fluctuations help with the hole's delocalization and, therefore, its mobility becomes larger for the same strength of pinning potential.

By comparing the Hubbard and \tJs\ models in Fig.~\ref{fig:3rdConnected}(c), we note that the results obtained from both models agree quantitatively at the small $V$ side, suggesting that the third-order spin-hole correlations originate from hole motion in an AFM background instead of any interactions beyond the $\mathcal{O}(t^2/U)$. 
On the large $V$ side, $B^{c}(\rbf, \rbf^\prime, \mathbf{d})$ displays slightly different temperature dependence for two models, which can be attributed to the modified exchange coupling in the vicinity of the pinning site: Between this site and its nearest neighbors, the superexchange coupling becomes $J_{\rm eff} = J  U^2 / (U^2-V^2)  > J $. At much higher temperatures ($T>J$), thermal fluctuations smear out the AFM spin correlations and thus suppress the third-order correlator.

\section{The Fifth-Order Ring-Spin Correlators}\label{sec:5thOrder}
\begin{figure}[!th]
\begin{center}
\includegraphics[width=\columnwidth]{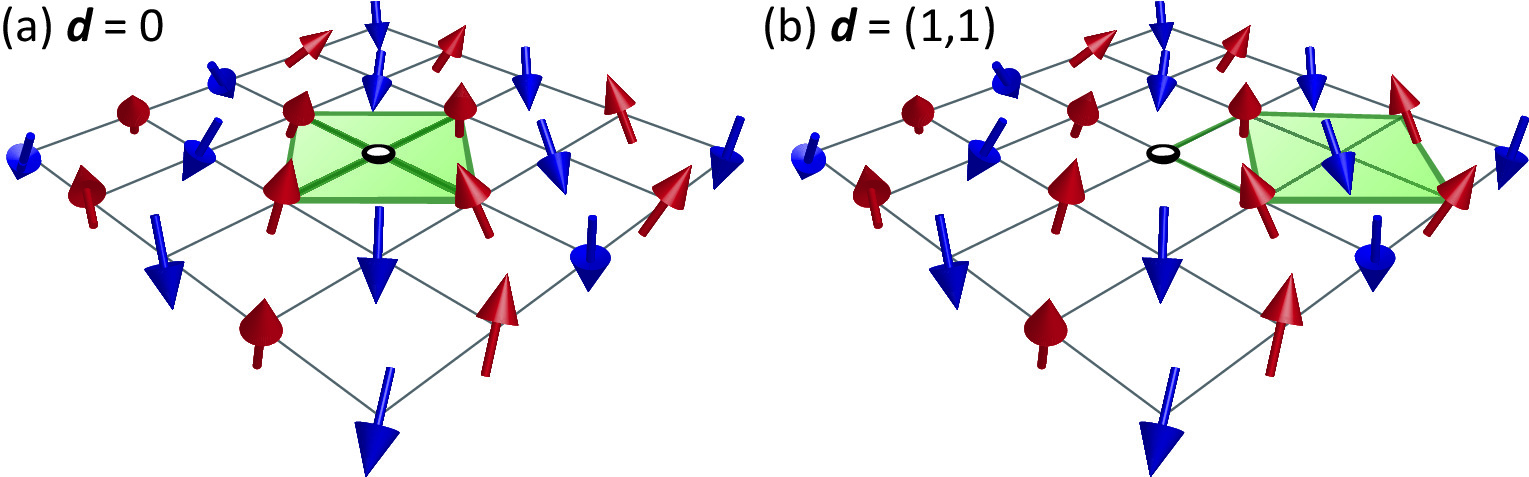}
\caption{\label{fig:5thCartoon} Schematic showing the fifth-order ring-spin correlators for (a) $\mathbf{d}=(0,0)$ and (b) $\mathbf{d}=(1,1)$.
}
\end{center}
\end{figure}

\begin{figure}[!t]
\begin{center}
\includegraphics[width=\columnwidth]{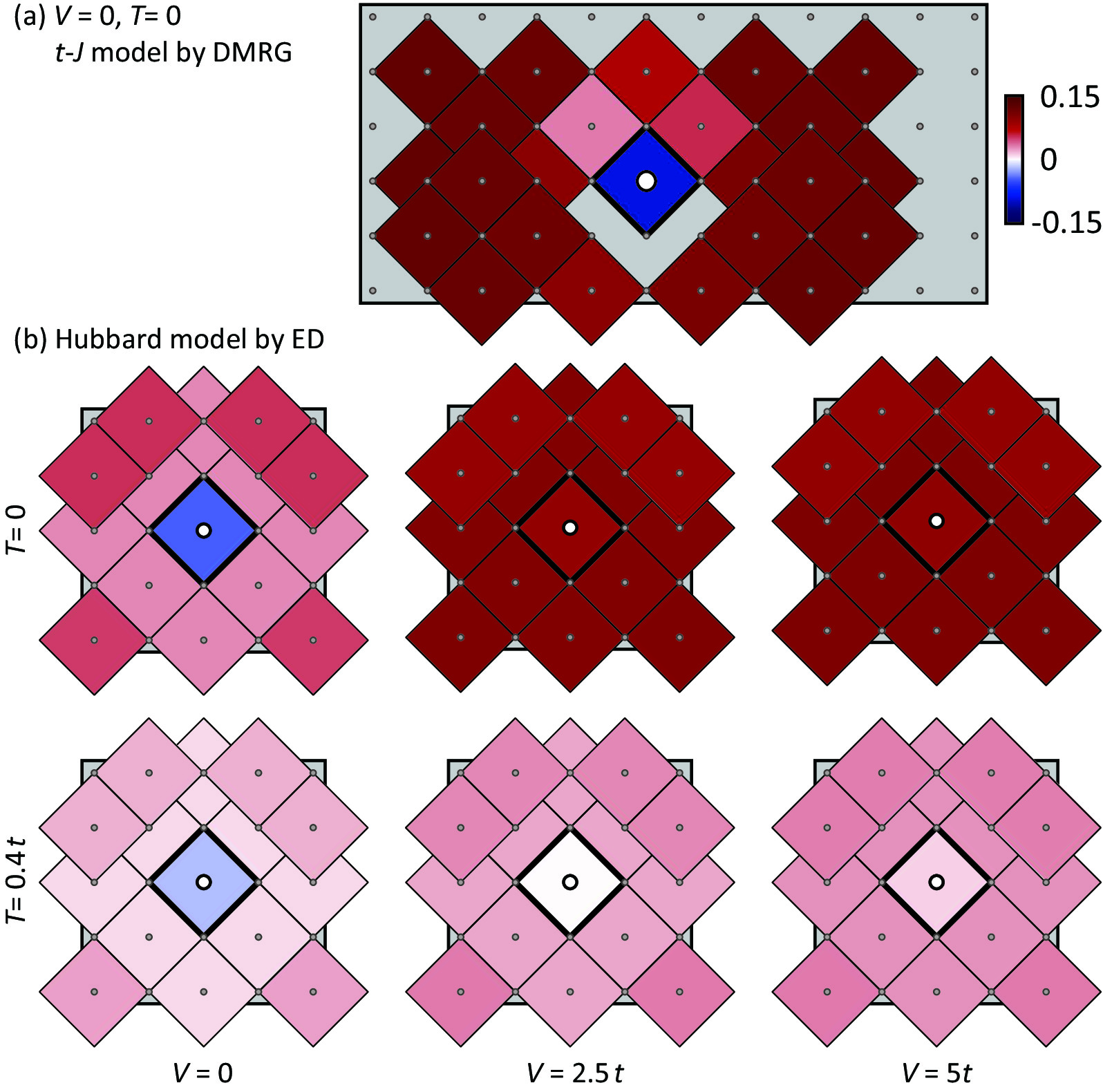}
\caption{\label{fig:ringSpinDistr} (a) The distribution of $C_\lozenge(\rbf, \textbf{d})$ for different distances $\textbf{d}$, $\rbf$ in the center and $V=0$, calculated for a 6$\times$12 \tJ\ model at zero temperature by DMRG. The weak asymmetry in the $x$-direction is caused by the non-central location of the hole at $x=6$, while the system is translational invariant along the $y$-direction due to the cylindrical boundary condition. (b) The distribution of $C_\lozenge(\rbf=\rbf_0, \textbf{d})$ for different distances $\textbf{d}$, strengths $V$ of the pinning potential in the center at $\rbf_0$, and temperatures $T$, calculated for the Hubbard model in a $4\times4$ cluster. The central white circle denotes the position of the dopant hole $\rbf_0$ on the site of the pinning potential. 
}
\end{center}
\end{figure}

The third-order correlators $B(\rbf, \rbf^\prime, \mathbf{d})$ reflect the spin fluctuations around the hole. These correlators are expected to be weak for mobile holes because the underlying motion of the dopants admixes spins from different sub-lattices in the surrounding AFM\cite{grusdt2018parton,grusdt2019microscopic}. To minimize the effect of averaging over different trajectories and provide deeper insights into the underlying spin-charge correlation, one may consider constructing higher-order correlators. 
To this end, we examine the (fifth-order) ring-spin correlators with respect to a doped hole, which is introduced recently in Ref.~\onlinecite{bohrdt2020shortPaper} as
\begin{equation}
C_\lozenge(\rbf, \textbf{d}) = 2^4\left\langle \hat{n}^{\rm h}_{\rbf} \hat{S}^z_{\rbf+\textbf{d}+\hat{x}} \hat{S}^z_{\rbf+\textbf{d}+\hat{y}} \hat{S}^z_{\rbf+\textbf{d}-\hat{x}} \hat{S}^z_{\rbf+\textbf{d}-\hat{y}} \right\rangle/\langle \hat{n}_{\bf r}^h\rangle.
\label{eqClozDef}
\end{equation}
As illustrated in Fig.~\ref{fig:5thCartoon}, this correlator reflects the ring-spin correlations in a co-moving frame of the doped hole. By including non-zero displacements $\bf d$ (between the ring-center and the hole), Eq.~\eqref{eqClozDef} generalizes the five-point correlators introduced in Ref.~\onlinecite{bohrdt2020shortPaper}.

\begin{figure*}[!th]
\begin{center}
\includegraphics[width=16cm]{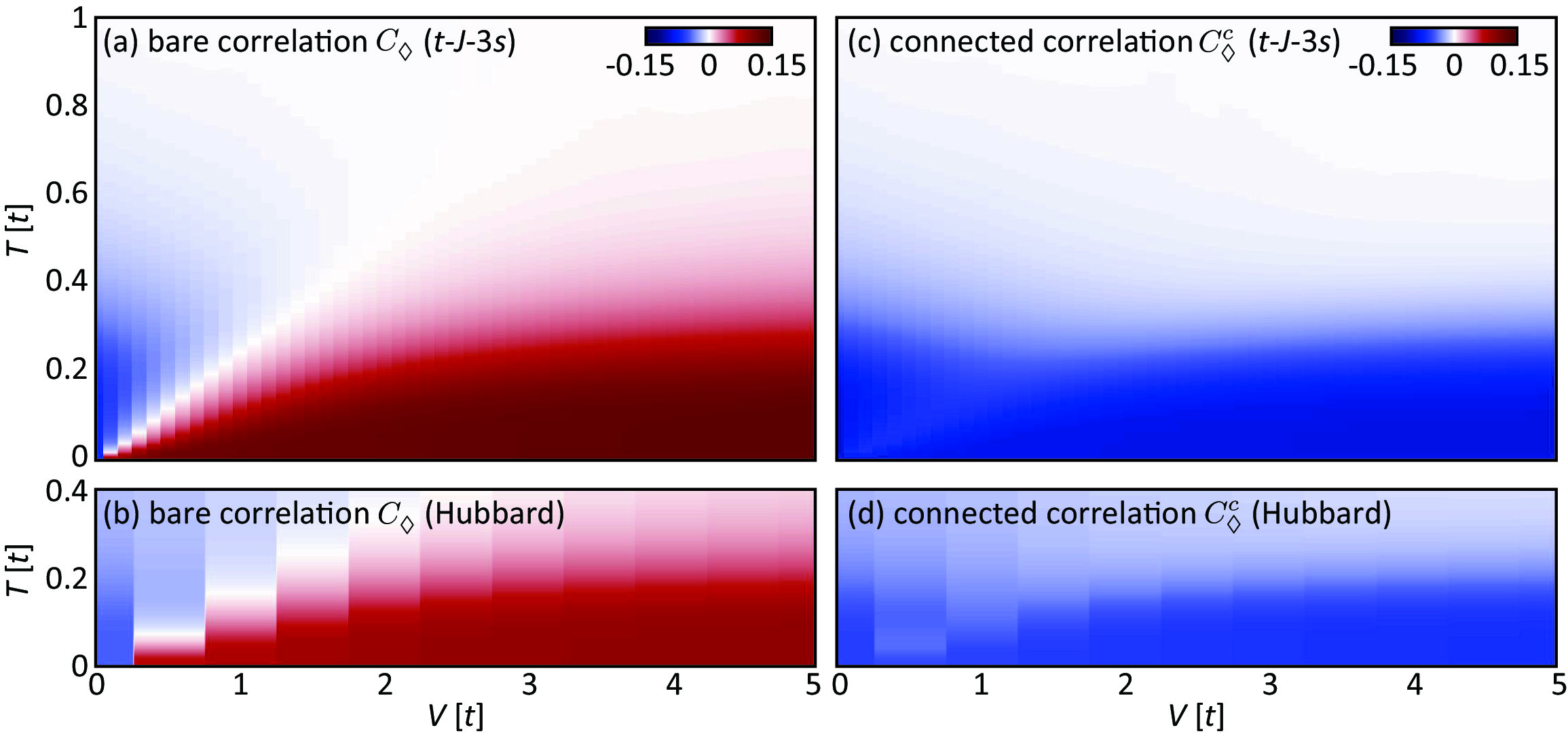}
\caption{\label{fig:ringSpinPhase} (a,b) The bare on-site ring-spin correlator $C_\lozenge(\rbf_0, \textbf{d}=\mathbf{0})$ as a function of pinning potential $V$ and temperature $T$, calculated using the (a) \tJs\ model and (b) Hubbard model. (c,d) Same as (a,b) but for the connected part of the on-site ring-spin correlator $C^c_\lozenge(\rbf_0, \textbf{d}=\mathbf{0})$ as a function of pinning potential $V$ and temperature $T$. 
}
\end{center}
\end{figure*}

We first investigate the dependence of $C_\lozenge(\rbf, \textbf{d})$ on the ring displacement $\textbf{d}$, and consider a mobile hole with no pinning potential ($V=0$). In this case, the system has translational invariance and $C_\lozenge(\rbf, \textbf{d})$ is only a function of $\textbf{d}$. As shown in Fig.~\ref{fig:ringSpinDistr} (a) and the top-left panel of (b), $C_\lozenge$ is positive for any $|\textbf{d}|>1$, but becomes sizable and negative for $\textbf{d}=\mathbf{0}$. This behavior is reminiscent of a $\mathbb{Z}_2$ Gauss law expected in a string description of the single spin polaron: when the mobile hole forms a spin polaron, its motion creates a ``string'' of displaced spins; In the $C_\lozenge$ correlator, spins which are part of the string originate from the opposite sub-lattice and contribute a negative sign\cite{bohrdt2020shortPaper}. Since $C_\lozenge(\rbf, \mathbf{0})$ involves only one site belonging to the string, it is expected to be negative; whereas $C_\lozenge(\rbf, {\bf d \neq \mathbf{0}})$ has statistically more contributions from an even number of spins affected by a string, displaying a positive value. We find they reach a quantitative agreement by comparing the \tJ\ model and the Hubbard model with different system sizes. This indicates that the observed correlations can be understood from the spin-exchange picture, while higher-order terms in the $t/U$ expansion can be ignored. The consistency between the DMRG calculation in a sizeable system and the ED calculation in a 4$\times$4 cluster further precludes the influence of system size and models.

We then consider systems with a finite pinning potential $V$ in the center at site $\rbf_0$. As shown in the upper panel of Fig.~\ref{fig:ringSpinDistr}(b), the on-site ($\bf d = \mathbf{0}$) correlator $C_\lozenge(\rbf_0, \mathbf{0})$ flips sign with a $V$-field. This is a direct consequence of a hole becoming immobile: when the pinning potential traps the dopant hole, the spin polaron breaks down, and the remaining system becomes a half-filled AFM system with a static defect. Since next-nearest spins are expected to be parallel in a pure N\'eel state, the $C_\lozenge(\rbf, \mathbf{0})$ becomes positive in this immobile case. This difference induced by the hole's mobility further reflects the screening to anti-screening crossover, consistent with the observation using the third-order correlators in Sec.~\ref{sec:3rdOrder}.

The sensitivity to a pinning potential is affected by the thermal fluctuation at finite temperature. As shown in the lower panel of Fig.~\ref{fig:ringSpinDistr}(b), the on-site correlator $C_\lozenge(\rbf_0, \mathbf{0})$ remains negative until $V\sim 2.5t$ at a higher temperature $T=0.4t$. As mentioned in Sec.~\ref{sec:3rdOrder}, this is a consequence of the hole's delocalization enhanced by thermal fluctuations.  To better visualize this crossover, we extract the $C_\lozenge(\rbf_0, \mathbf{0})$ and plot its $T$ and $V$ dependence in Figs.~\ref{fig:ringSpinPhase}(a) and (b). As temperatures rise, the regime dominated by the spin-polaron screening (blue) extends to a larger pinning potential. We observe a similar crossover here, as in Fig.~\ref{fig:3rdConnected}; note that the crossover's exact position may be different for different observables as it is not a phase transition. Except for this crossover, we find that all $C_\lozenge(\rbf_0, \textbf{d})$s decrease with increasing temperature, which occurs dramatically near $T\sim J$ and indicates the thermal melting of the surrounding antiferromagnet. Given the presence of four spins in the fifth-order correlator, its temperature dependence is more evident than the third-order one presented in Fig.~\ref{fig:3rdConnected}(c).

As such, the temperature dependence of this fifth-order correlator reflects the screening to anti-screening crossover similar to the third-order one. The difference is that this phenomenon has already been reflected in the ``bare'' fifth-order correlator, without spatial derivative or subtracting disconnected parts. In contrast, the genuine (connected) fifth-order correlator reflects a different level of information. To extract the genuine fifth-order correlators, we subtract off the disconnected pieces of the correlator following the same manner as Ref.~\onlinecite{bohrdt2020shortPaper}:
\begin{widetext}
\begin{multline}
C^{c}_\lozenge(\rbf,\textbf{d}) = C_\lozenge(\rbf,\textbf{d}) - 2^4  \biggl[ \sum_{\mathbf{l} \notin (\mathbf{i},\mathbf{j},\mathbf{k})} \hspace{-2.5mm}
\frac{ \langle \hat{n}^{\rm h}_{\mathbf{r}} \hat{S}^z_{\rbf +\textbf{d}+ \mathbf{i}} \hat{S}^z_{\rbf +\textbf{d}+ \mathbf{j}} \hat{S}^z_{\rbf +\textbf{d}+ \mathbf{k}}\rangle_c}{\langle \hat{n}^{\rm h}_{\mathbf{r}}\rangle} 
 \frac{\langle \hat{n}^{\rm h}_{\rbf} \hat{S}^z_{\rbf +\textbf{d}+ \mathbf{l}} \rangle}{{\langle \hat{n}^{\rm h}_{\mathbf{r}}\rangle}}
+  \hspace{-2.5mm} \sum_{(\mathbf{i}, \mathbf{j}) \notin (\mathbf{k}, \mathbf{l})} 
  \hspace{-3.5mm} \frac{\langle \hat{n}^{\rm h}_{\rbf} \hat{S}^z_{\rbf +\textbf{d}+ \mathbf{i}} \hat{S}^z_{\rbf +\textbf{d}+ \mathbf{j}} \rangle_c}{\langle \hat{n}^{\rm h}_{\mathbf{r}}\rangle} 
 \frac{\langle \hat{n}^{\rm h}_{\rbf} \hat{S}^z_{\rbf +\textbf{d}+ \mathbf{k}} \hat{S}^z_{\rbf +\textbf{d}+ \mathbf{l}} \rangle_c}{\langle \hat{n}^{\rm h}_{\mathbf{r}}\rangle} \\
+ \sum_{\mathbf{i} \neq \mathbf{j} \notin (\mathbf{k}, \mathbf{l})}  
 \frac{ \langle \hat{n}^{\rm h}_{\rbf} \hat{S}^z_{\rbf +\textbf{d}+ \mathbf{i}} \rangle }{\langle \hat{n}^{\rm h}_{\mathbf{r}}\rangle}
 \frac{ \langle \hat{n}^{\rm h}_{\rbf} \hat{S}^z_{\rbf +\textbf{d}+ \mathbf{j}} \rangle }{\langle \hat{n}^{\rm h}_{\mathbf{r}}\rangle}
 \frac{ \langle  \hat{n}^{\rm h}_{\rbf} \hat{S}^z_{\rbf +\textbf{d}+ \mathbf{k}} \hat{S}^z_{\rbf +\textbf{d}+ \mathbf{l}} \rangle_c }{\langle \hat{n}^{\rm h}_{\mathbf{r}}\rangle} 
+ \frac{ \langle  \hat{n}^{\rm h}_{\rbf}  \hat{S}^z_{\rbf+\textbf{d}+\hat{x}} \rangle }{\langle \hat{n}^{\rm h}_{\mathbf{r}}\rangle} 
  \frac{ \langle  \hat{n}^{\rm h}_{\rbf}  \hat{S}^z_{\rbf+\textbf{d}+\hat{y}} \rangle  }{\langle \hat{n}^{\rm h}_{\mathbf{r}}\rangle}
  \frac{ \langle  \hat{n}^{\rm h}_{\rbf}  \hat{S}^z_{\rbf+\textbf{d}-\hat{x}} \rangle  }{\langle \hat{n}^{\rm h}_{\mathbf{r}}\rangle}
  \frac{ \langle  \hat{n}^{\rm h}_{\rbf}  \hat{S}^z_{\rbf+\textbf{d}-\hat{y}} \rangle }{\langle \hat{n}^{\rm h}_{\mathbf{r}}\rangle}
\biggr],
\label{eqCconnDef}
\end{multline}
\end{widetext}
where the $\langle \cdots\rangle_c$ denotes the connected correlator with lower-order terms subtracted.

Figures~\ref{fig:ringSpinPhase}(c,d) show the dependence of the connected correlator $C^c_\lozenge(\rbf,\mathbf{0})$ on the temperature and pinning potential. We find that this correlator is sizably negative even for large pinning potentials. Although the characteristic temperature, above which $|C^c_\lozenge|$ drops dramatically, decreases slightly for the systems with pinned holes, the absolute value of $C^c_\lozenge$ does not change much with $V$. This observation indicates that the genuine fifth-order correlator is always present, significantly deviating from the classical N\'eel state regardless of whether the hole is mobile. Such an intrinsic nature is invisible in lower-order correlators.

In the remainder of this section, we elucidate the physical origin and properties of the genuine fifth-order ring-spin correlators. We discuss the role of different statistical ensembles on the fifth-order ring-spin correlators. 

To this end, we compare a spin-imbalanced ensemble with definite total spin $S^z=1/2$ to a spin-balanced statistical mixture with $\langle \hat{S}^z \rangle =0$. Alternatively, in the former case, we could consider an ensemble with definite $S^z=-1/2$, which gives the same results: the higher-order correlators $C_\lozenge$ and $C_\lozenge^c$ each involve an even number of spin operators $\hat{S}^z$. Moreover, the underlying spin or Hubbard Hamiltonian is invariant under a global spin-flip ($S^z_j \to - S^z_j$). Hence, the correlators in the sectors with definite $S^z=\pm 1/2$ are equivalent, 
\begin{equation}
C_\lozenge |_{S^z=1/2} = C_\lozenge |_{S^z=-1/2}, \quad C_\lozenge^c |_{S^z=1/2} = C_\lozenge^c |_{S^z=-1/2}.
\end{equation}
With the same argument, in the spin-balanced ensemble, $\langle \hat{S}^z \rangle =0$, all lower-order correlators involving an odd number of spin operators vanish. 
While an ensemble with definite $S^z=1/2$ or $-1/2$ is hard to realize in a solid-state system, both spin ensembles can be addressed experimentally with ultracold atoms and with full spin and charge resolution\cite{boll2016spin,koepsell2020robust}. This ensemble can be achieved by post-selecting experimental data with specific $S^z$ values.

The two spin ensembles correspond to different ways of taking the zero-temperature limit. The usual canonical ensemble converges to a balanced mixture of the two sectors $S^z=+1/2$ and $S^z=-1/2$ when $T \to 0$. In the resulting spin-balanced ensemble $\langle \hat{S}^z \rangle=0$, and all but the second sum on the right-hand side of Eq.~\eqref{eqCconnDef} vanish. In contrast, for the ground state with a definite $S^z=1/2$, all terms in Eq.~\eqref{eqCconnDef} contribute.

\begin{figure}[!t]
\begin{center}
\includegraphics[width=\columnwidth]{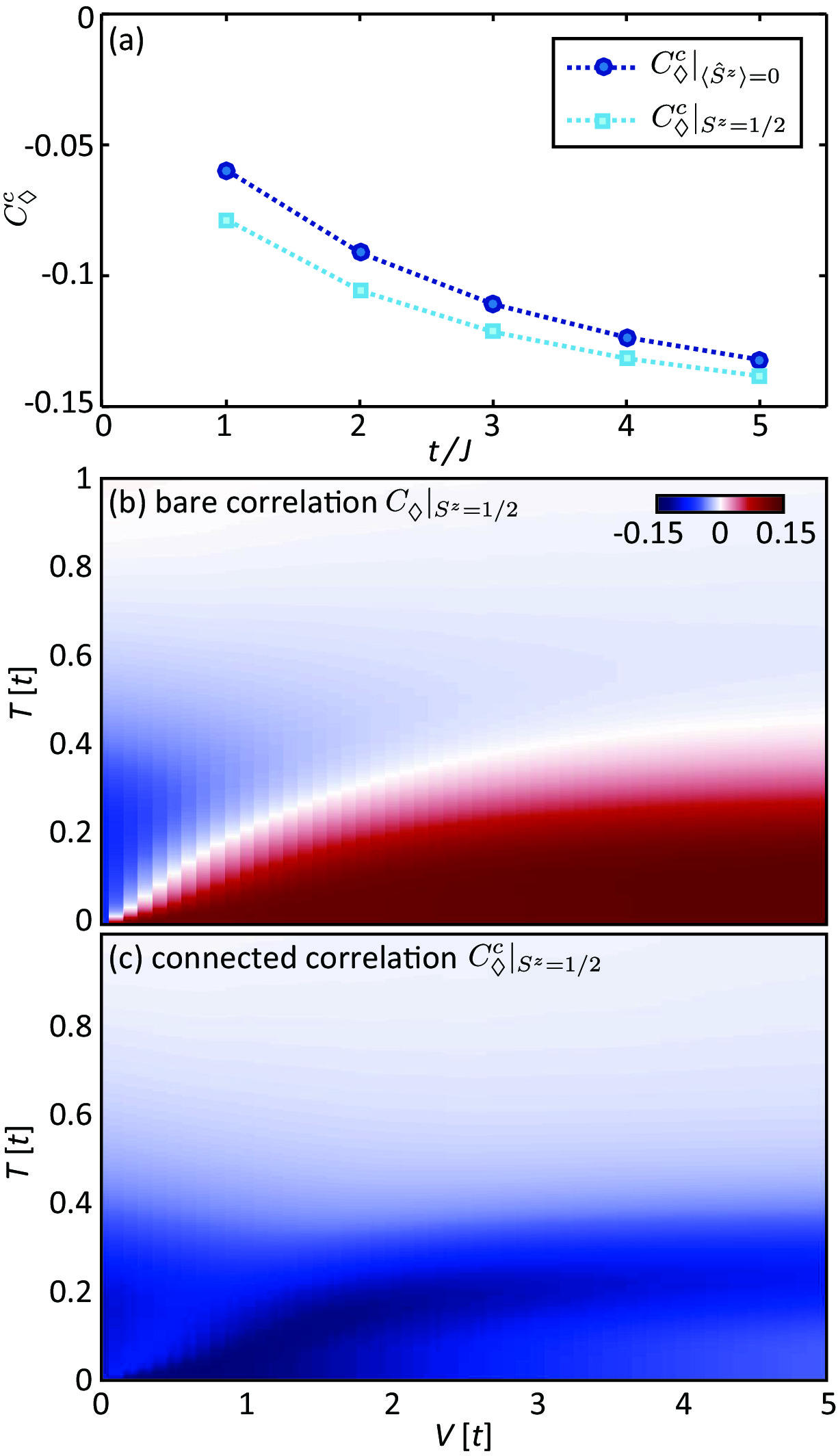}
\caption[The fifth-order correlators evaluated from a spin-imbalanced ensemble]{\label{fig:ringSpinPhaseSz12} (a) Genuine fifth-order correlator $C^c_\lozenge$ for different spin-ensembles, $\langle \hat{S}^z \rangle = 0$ and $S^z=\pm 1/2$, at zero temperature and without pinning potential. The results are obtained by the DMRG simulations for a \tJ\ Hamiltonian on a $6\times 12$ cylinder. (b) The bare fifth-order correlator $C_\lozenge$ and (c) the genuine correlator $C^c_\lozenge$ in a \emph{spin-imbalanced} ensemble with fixed total $S^z=1/2$, as a function of pinning potential $V$ and temperature $T$. We consider a \tJs\ Hamiltonian.
}
\end{center}
\end{figure}

Nevertheless, for a mobile hole we find that the calculated connected correlators do not change significantly if switched to the imbalanced ensemble: In Fig.~\ref{fig:ringSpinPhaseSz12}(a), we plot $C_\lozenge^c |_{\langle S^z \rangle = 0}$ as a function of $t/J$ (see Fig.~1 in \onlinecite{bohrdt2020shortPaper}) and compare it to the spin imbalanced ensemble. Qualitatively, the same behavior is found for all considered values of $t/J$. The deviations are largest for small values of $t/J$, where we expect a smaller spin-polaron radius in the ground state \cite{grusdt2019microscopic}.

The situation becomes slightly different when the hole is trapped by a pinning potential. In Fig.~\ref{fig:ringSpinPhaseSz12} we compare connected and bare correlators for a \emph{spin-imbalanced} ensemble with the fixed total $S^z=1/2$. The bare correlator $C_\lozenge$ in Fig.~\ref{fig:ringSpinPhaseSz12}(b) reflects a crossover similar to Fig.~\ref{fig:ringSpinPhase}(a), with slightly stronger temperature dependence. This sensitivity to temperature results from the fact that the fixed total spin $S^z=1/2$ gives finite contributions to the odd terms in Eq.~\eqref{eqCconnDef} as lower orders and thus induces more fluctuations when temperature increases. For the same reason, the connected fifth-order correlator $C^c_\lozenge$ also displays a slightly stronger sensitivity to the pinning potential $V$s [see Fig.~\ref{fig:ringSpinPhaseSz12} (b)].

Most significantly, we observe in Fig.~\ref{fig:ringSpinPhaseSz12}(c) that the connected correlator $C^c_\lozenge$ decreases in magnitude in the spin-imbalanced ensemble for large $V$ and small temperatures $T$; in contrast, it remains sizably negative in this regime for the spin-balanced ensemble [see Fig.~\ref{fig:ringSpinPhase}(c)]. We attribute this influence of the spin ensemble on the connected higher-order correlator to the formation of N\'eel order and a non-zero staggered magnetization, $\langle \hat{M}^z \rangle = \sum_{\rbf} (-1)^{r_x+r_y} \langle \hat{S}^z_{\rbf} \rangle \neq 0$, in the lattice. In a finite-size system, like the one we study, the latter requires spin-imbalance and a pinning potential for the hole to avoid mixing of A- and B- sublattices in the co-moving frame of the hole where we define the higher-order correlators. When a non-zero staggered magnetization forms in the co-moving frame with the hole, the lower-order terms -- e.g. $\langle \hat{S}^z_{\rbf} \rangle$ -- provide sizable non-vanishing contributions to the bare ring-spin correlators. Since we observe that bare ring-spin correlators are only weakly affected by different spin ensembles, the genuine higher-order terms can differ significantly for different ensembles, as we find in Fig.~\ref{fig:ringSpinPhaseSz12}(b).

Finally, to further illustrate the differences between spin-balanced and imbalanced ensembles, we consider a strongly simplified physical setting at zero temperature. We assume that the hole is fully pinned in the center of the system ($V \gg t$) and take into account only Ising couplings $\sim J^z \hat{S}^z_{\mathbf{i}} \hat{S}^z_{\mathbf{j}}$ between the spins. The ground state of this toy model in the spin-imbalanced ensemble with $S^z=1/2$ is a classical N\'eel state $|{\rm N},+\rangle$ where a missing down-spin at $\rbf$ defines the pinned hole. Similarly, the ground state in the spin-imbalanced ensemble with $S^z=-1/2$ is the opposite N\'eel state $|{\rm N},-\rangle$, obtained from $|{\rm N},+\rangle$ by flipping all spins.
For both N\'eel states, the bare ring-spin correlator $C_\lozenge(\rbf,\mathbf{0})|_{S^z=1/2} = C_\lozenge(\rbf,\mathbf{0})|_{S^z=-1/2}$ is equal since all four spins surrounding the hole are aligned. Hence, $C_\lozenge(\rbf,\mathbf{0})|_{\langle \hat{S}^z \rangle=0} = C_\lozenge(\rbf,\mathbf{0})|_{S^z=\pm 1/2}$ also takes the same value in the spin-balanced ensemble, defined by the even statistical mixture $\{ |{\rm N},+\rangle, |{\rm N},-\rangle \}$. Now we turn to the connected correlators. For each of the two N\'eel states $|{\rm N},\pm\rangle$ taken individually,  $C_\lozenge^{c}|_{S^z=\pm 1/2} = 0$ vanishes: By construction, the higher-order connected correlators are identically zero for product states. To obtain this result from Eq.~\eqref{eqCconnDef}, it is important to account for all lower-order contributions, including those with an odd number of $\hat{S}^z$ operators. In contrast, in the spin-balanced ensemble where $\langle \hat{S}^z \rangle = 0$, only lower-order terms with an even number of $\hat{S}^z$ operators contribute in Eq.~\eqref{eqCconnDef}. This leads to a strongly modified connected correlator, with an overall negative value: indeed, counting all terms for the classical N\'eel states gives the estimate $C^c_\lozenge(\rbf,\mathbf{0})|_{\langle \hat{S}^z \rangle=0} = -2 $ in our toy model with Ising interactions.

\section{Discussions and Conclusions}\label{sec:conclusion}
Using the third-order and fifth-order spin-hole correlators, we have analyzed spatial correlation of a doped hole and its surrounding spin fluctuations in the single-hole doped Hubbard and \tJs\ models. We particularly investigate the impact of impurity on these high-order correlations, mimicked by a localized pinning potential with varying strength $V$. Interestingly, we find that even an extremely local impurity imposes an outsized effect on the high-order correlations. With the increase of the pinning potential, we identified a crossover from screening to anti-screening of the spin fluctuations surrounding the hole: For a mobile hole, the weakened third-order correlators in proximity to the dopant and the negative fifth-order ring-spin correlators reflect the underlying spin polaron formation -- here the motion of the hole screens the spin defects; for an immobile hole trapped by the pinning potential, the third-order correlators in proximity to the dopant are enhanced and the bare fifth-order ring-spin correlators are positive -- as a result of the anti-screened magnetic moment of the pinned hole. In this case, a geometric defect and a staggered field are induced in the AFM background. 
This microscopic transition from a spin-polaron to an anti-screened defect provides a route to understand the Zn-substitution experiments, where the magnetic moment is enhanced while superconductivity is suppressed. Due to the breaking of the spin-polaron, different holes are no longer paired by sharing the spin fluctuations. Instead, the locally enhanced spin fluctuations surrounding, the impurity-trapped hole, indicates a repulsion to other holes and therefore suppresses the coherence of Cooper pairs. This crossover in high-order correlators may also provide intuitions to impurity effects in other correlated materials.

We have also examined the impact of temperature on the considered spin-charge correlations. We find that thermal fluctuations lead to an extended spin-polaron regime with a weakly pinned hole, and identify the crossover to the pinned hole as a function of both temperature $T$ and pinning potential $V$. 
In contrast to the crossover, we found genuine (i.e. connected) fifth-order ring-spin correlators carrying different underlying physics: they are sizably negative and robust against the pinning potential. 

Based on these observations, we demonstrated that higher-order correlation functions provide a new perspective on quantum many-body systems with strong correlations. This is a particularly promising direction in the context of correlated fermionic systems such as the Fermi-Hubbard model, where traditional theoretical approaches are limited and a comprehensive physical picture is still lacking. The intrinsic properties of this type of quantum many-body systems may be invisible in traditional two-point correlations of solid-state measurements.  The specific situation discussed here, where we analyzed the effect of a localized pinning potential on mobile dopants, is particularly relevant in the solid-state context for understanding the effects of disorder on strong spin-charge correlations. On the other hand, the higher-order correlators we propose to measure are directly accessible in state-of-the-art quantum gas microscopy experiments with ultracold atoms in optical lattices. The required temperatures to observe non-trivial effects have already been reached, and we expect that even lower temperatures will become accessible in the near future.

\section*{Acknowledgments}
Y.W. acknowledges support from the National Science Foundation (NSF) award DMR-2038011.
A.B. and F.G. acknowledge support from Deutsche Forschungsgemeinschaft (DFG, German Research Foundation) under Germany’s Excellence Strategy – EXC-2111 – 390814868. J.K. acknowledges funding from Hector Fellow Academy and support from the Max Planck Harvard Research Center for Quantum Optics (MPHQ). E.D. acknowledges support from Harvard-MIT CUA, ARO grant number W911NF-20-1-0163, the NSF award OAC-193471, and Harvard Quantum Initiative. This research used resources of the National Energy Research Scientific Computing Center (NERSC), a U.S. Department of Energy Office of Science User Facility operated under Contract No. DE-AC02- 05CH11231.

\appendix
\begin{figure*}[!th]
\begin{center}
\includegraphics[width=12cm]{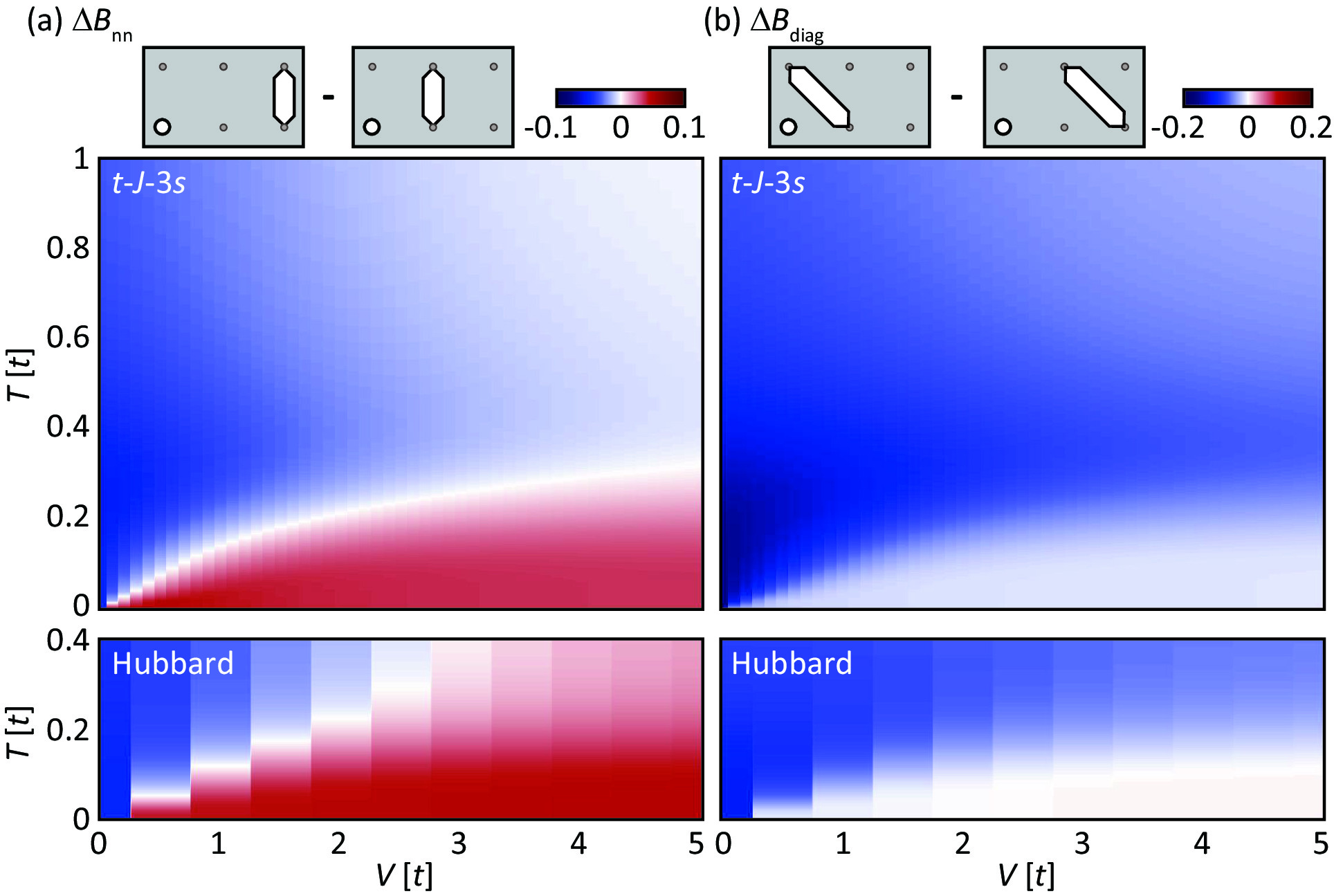}
\caption{\label{fig:3rdPhaseDiagDifferential} (a) Dependence of the differential third-order correlators $\Delta B_{\rm nn}$ with bond lengths $|\mathbf{d}|=1$ as a function of the trapping potential $V$ and temperature $T$, obtained for the \tJs\ model (top) and the Hubbard model (bottom).
(b) The same as (a) but for $\Delta B_{\rm diag}$ with $|\mathbf{d}|=\sqrt{2}$. The insets in both (a) and (b) explain the definition of the differential correlators. 
}
\end{center}
\end{figure*}
\section{Details about the Differential Third-Order Correlator}

The differential third-order correlator has been briefly discussed in Sec.~\ref{sec:3rdOrder}. To reveal the impact of temperature and the pinning potential, we present its $V$-$T$ dependence of $\Delta B_{\rm nn}$ in Fig.~\ref{fig:3rdPhaseDiagDifferential}, following the same manner of Fig.~\ref{fig:3rdConnected}(c) in the main text.

In addition to the $\Delta B_{\rm nn}$ defined in Eq.~\ref{eq:3rdDiffNN}, we further quantify the difference between the diagonal (hole-spin) correlators as 
\begin{equation}
    \Delta B_{\rm diag} = B(\rbf, \rbf +\hat{x}, \hat{y}-\hat{x}) - B(\rbf, \rbf +2 \hat{x}, \hat{y}-\hat{x}) \,.
\end{equation}
The spatial relations of both differences are illustrated in the inset of Fig.~\ref{fig:3rdPhaseDiagDifferential}. Note, to compensate for the negative sign induced by the AFM sublattices, we swapped the order while defining these two differences.

We notice a difference between the $\Delta B_{\rm nn}$ [Fig.~\ref{fig:3rdPhaseDiagDifferential}(a)] and the $\Delta B_{\rm diag}$ [Fig.~\ref{fig:3rdPhaseDiagDifferential}(b)]: although the screening effect has been dramatically suppressed for large pinning potentials following the similar crossover, $\Delta B_{\rm diag}$ does not exhibit an evident sign change. Since spin correlations are inhomogeneous when the system breaks translational symmetry regardless of the hole, the differential correlator $\Delta B_{\rm diag}$ has been affected by this background inhomogeneity, reducing the signal-to-noise ratio for describing the underlying properties of the spin-hole composite.

The advantage of the genuine correlator $B^c(\rbf, \rbf^\prime, \mathbf{d})$ over the differential correlator $\Delta B_{\rm diag}$ is reflected by comparing Fig.~\ref{fig:3rdConnected}(c) and Fig.~\ref{fig:3rdPhaseDiagDifferential}(b).
Although both describe the diagonal spin correlations relative to the hole, the genuine correlator clearly separates the spin polaron (where it is sizeably negative) and the anti-screening regime (where it is sizeably positive). Especially in the large $V$ limit, the genuine correlator characterizes a narrower crossover compared with Fig.~\ref{fig:3rdConnected}(a), due to the exclusion of noise in lower-order correlators.

\bibliography{paper_cold_atom}

\begin{thebibliography}{118}
\bibitem[{deg()}]{degeneracy}
\bibinfo{note}{The fact that the correlation switches sign for any
  infinitesimal $V$ results from the extra hypercubic symmetry in a $4\times4$
  Hubbard model. This symmetry leads to a 6-fold degeneracy for the ground
  state without $V$, and is broken for any finite $V$. The critical $V$ should
  lie at a small finite value at the thermodynamic limit, similar to the
  presented results at finite temperature.}
\end{thebibliography}

\end{document}